\documentclass[a4paper,preprintnumbers,nofootinbib,noshowpacs,prd]{revtex4}

\usepackage{graphicx,epsfig}
\usepackage{amsmath,amssymb,bbold}
\usepackage{url}
\usepackage{rotating}
\usepackage{color}
\usepackage{hhline}
\usepackage[pass]{geometry}
\usepackage{ulem,slashed}

\graphicspath{ {Plots/} }

\makeatletter       

\renewcommand{\p@subsection}{}

\makeatother

\newcommand{\onehalf}{\ensuremath{\frac{1}{2}}}

\begin{document}

\begin{flushright}
PITT-PACC-1504 \\
\end{flushright}

\vskip 0.5cm

\title{Characterizing Invisible Electroweak Particles through \\
       Single-Photon Processes at High Energy \boldmath{$e^+e^-$} Colliders}

\author{S.Y. Choi$^{1,2}$, Tao Han$^{2,3,4}$, J. Kalinowski$^{5}$,
        K. Rolbiecki$^{5,6}$ and Xing Wang$^{2}$ \\[-3mm]
        \mbox{ }\\
   $^1$ {\it Department of Physics, Chonbuk National University, Jeonbuk 561-756,
             Korea} \\
   $^2$ {\it Pittsburgh Particle Physics, Astrophysics, and Cosmology
             Center, Department of Physics and Astronomy, University of Pittsburgh,
             Pittsburgh, PA 15260, USA}\\
   $^3$ {\it Center for High Energy Physics, Tsinghua University, Beijing 100084,
             China} \\
   $^4$ {\it Korea Institute for Advanced Study (KIAS), Seoul 130-012, Korea}\\
   $^5$ {\it Faculty of Physics, University of Warsaw, 02093 Warsaw, Poland}\\
   $^6$ {\it IFT-UAM/CSIC, C/ Nicol\'as Cabrera 13-15, 28049 Madrid, Spain}}

\date{\today}

\begin{abstract}
\noindent
We explore the scenarios where the only accessible new states
at the electroweak scale consist of a pair of color-singlet electroweak
particles, whose masses are degenerate at the tree level and split only by
electroweak symmetry breaking at the loop level. For the sake of illustration,
we consider a supersymmetric model and study the following three
representative cases with the lower-lying states as (a) two spin-1/2
Higgsino SU(2)$_L$ doublets,
(b) a spin-1/2 wino SU(2)$_L$ triplet and (c) a spin-0 left-handed slepton
SU(2)$_L$ doublet. Due to the mass-degeneracy, those lower-lying electroweak
states are difficult to observe at the LHC and rather challenging to detect at
the $e^+ e^-$ collider as well. We exploit the pair production in association
with a hard photon radiation in high energy $e^+ e^-$ collisions.
If kinematically accessible, such single-photon processes at $e^+e^-$ colliders
with polarized beams enable us to characterize each scenario by measuring the
energy and scattering angle
of the associated hard photon, and to determine the spin of the nearly invisible
particles unambiguously through the threshold behavior in the photon energy distribution.
\end{abstract}

\maketitle


\section{Introduction}
\label{sec:introduction}

The discovery of the Higgs boson at the CERN Large Hadron
Collider (LHC) \cite{Aad:2012tfa,Chatrchyan:2012ufa} truly sets a milestone in
particle physics. It completes the structure of the standard model (SM), which
may be valid as a self-consistent effective theory all the way up to the Planck
scale. The rather light mass of 125 GeV \cite{mass125} and
narrow width of much less than a GeV \cite{Caola:2013yja,Khachatryan:2014iha}
for the Higgs boson imply a weakly coupled theory at work for the electroweak
symmetry breaking (EWSB) sector. Naturalness argument \cite{Wilson:1970ag,
Gildener:1976ai,Weinberg:1978ym,'tHooft:1979bh} thus prefers the existence
of new states associated with the EWSB sector.
Supersymmetry \cite{Nilles:1983ge,Haber:1984rc} is arguably the best motivated
candidate for a natural theory, and the relevant partners include
top squarks, gluinos, electroweak (EW) gauginos and
Higgsinos. Another important feature of supersymmetric (SUSY) models is the
lightest neutral SUSY particle (LSP) to  serve as a cold dark
matter (DM) candidate \cite{Griest:2000kj,Bertone:2004pz}.
However, it has been quite puzzling that except for a SM-like Higgs boson,
no new particles beyond the SM have been so far observed in the LHC experiments
near and above the TeV threshold. One plausible scenario for the LHC null search
results is that all the colored SUSY particles with QCD strong interactions are
rather heavy and thus out of reach \cite{ArkaniHamed:2004yi,Giudice:2004tc,
Wells:2004di,Baer:2012uy,Papucci:2011wy,Jung:2013zya}. The EW particles,
although kinematically accessible, may not lead to experimentally tractable signals
due to the rather small production rate, the un-characteristic signature and the
large SM backgrounds at hadron colliders \cite{Chen:1995yu,Chen:1996ap,
Giudice:2010wb,Han:2013kza,Cheung:2005pv,Baer:2011ec,Baer:2012uy,Hensel:2002bu,
Berggren:2013vfa,Cheng:1998hc,Gherghetta:1999sw,Cheung:2005ba,Feng:1999fu,
Ibe:2006de,Dimopoulos:1990kc,Thomas:1998wy,Sher:1995tc}.
This situation happens quite naturally when the lower-lying EW states
are nearly degenerate in mass, and thus the final state products are rather soft
and have little missing transverse energy.
On the other hand, the future $e^+e^-$ colliders, such as the International Linear
Collider (ILC) \cite{Behnke:2013xla,Baer:2013cma,Behnke:2013lya},
would be capable of covering the search as long as kinematically accessible,
because of the well-constrained event topology and the very clean experimental
environment.

In this paper, we set out to study this challenging scenario at an $e^+e^-$ collider
in a rather model-independent way, to quantify the observability for the missing
particle signal, and to explore the feasibility to determine the missing particle
mass, spin and chiral couplings.
Within a generic framework of the minimal supersymmetric standard model (MSSM),
we focus on three representative cases to study the EW lower-lying states,
where the other SUSY particles are assumed to be decoupled.
The first scenario, to be called the spin-1/2 Higgsino scenario,
is the case where the only accessible SUSY particles are two spin-1/2
Higgsino
doublets $(\tilde{H}^+, \tilde{H}^0)$. The second scenario, to be called the
spin-1/2 wino scenario, is the case where the only accessible SUSY particles are
a spin-1/2 wino triplet $(\tilde{W}^+, \tilde{W}^0, \tilde{W}^-)$.
The third scenario, to be called the spin-0 slepton scenario, is the case where
the only accessible SUSY particles consist of a spin-0 left-handed slepton
doublet $(\tilde{\nu}_\ell, \tilde{\ell}^-)$.

In each scenario, the charged particle and its neutral partner are degenerate
in mass before EWSB and their mass splitting originates dominantly from loop-induced
EWSB corrections in the Higgsino and wino scenarios, or from the so-called $D$-term
potential after EWSB in the slepton scenario.
Due to the near degeneracy it would be very challenging to observe the soft final
state particles.
Analogous to the mono-jet plus missing energy signature at hadron
colliders \cite{Chatrchyan:2012me,ATLAS:2012ky},
single energetic photon plus missing energy at $e^+e^-$ colliders is known to
be one of the promising search channels for the missing
particles \cite{Chen:1995yu,Hensel:2002bu,Berggren:2013vfa,Birkedal:2004xn}.
This method was used for counting neutrino families \cite{Ma:1978zm,
Gaemers:1978fe,Barbiellini:1981zm} and as a means to search for (nearly)
invisible SUSY particles \cite{Grassie:1983kq,Fayet:1986zc,Dicus:1990vm,
Lopez:1996gd,Dreiner:2006sb,Choi:1999bs,Bartels:2012ex}.
We provide systematic and detailed methods not only for determining the masses
and spins of the (nearly) invisible particles unambiguously, but also for
characterizing each of the three benchmark scenarios through single-photon
processes at $e^+e^-$ colliders by exploiting electron and positron beam
polarizations. We find that, if kinematically accessible, the
masses, spins and coupling structures of the invisible particles in
such single-photon processes can be determined clearly by exploiting the
initial electron (and positron) beam polarization and investigating the
threshold excitation patterns of the processes.

The remainder of the paper is organized as follows. We first set up the
three benchmark scenarios in the MSSM framework. We lay out their
spectra and interactions with the SM particles.
We present the mass splitting in each
scenario by radiative corrections or by $D$-term.
Section~\ref{sec:single-photon_processes} is devoted to systematic analyses
for the radiative processes involving the pair production and an associated
hard photon in $e^+e^-$ collisions with special emphasis on the comparison
of the initial-state radiation (ISR) and final-state radiation (FSR) in the
charged pair production. We present the dependence of the cross sections on
the photon energy and the electron/positron beam polarizations.
In Sec.$\,$\ref{sec:characteristics}, we first study the discovery
limit of the new invisible particles based on the statistical significance of
each mode at a 500 GeV ILC.
We then describe systematically how the threshold behavior and the ratios
of polarized cross sections enable us to determine the SUSY particle spin
and characterize each scenario unambiguously. We briefly comment on the other
alternative methods for characterizing the properties of the scenarios.
Finally, we summarize our results and present our conclusions
in Sec.$\,$\ref{sec:summary_conclusions}.

\section{Scenarios with a degenerate pair of SUSY particles}
\label{sec:scenarios}

To study the nearly degenerate EW states in
a relatively model-independent way, we take the MSSM as a generic framework and
make the following simple assumptions:
only a pair of SUSY color-singlet EW particles is kinematically accessible below
the ILC threshold, and the other heavier states are essentially decoupled.
This could be realized when the soft SUSY-breaking scalar quark masses
and the gluino mass scale $M_3$ are much heavier than the EW soft SUSY-breaking
scales.
Specifically, we consider three benchmark scenarios in MSSM, each representing
a qualitative different case, as described in detail below.

\subsection{The spin-1/2 Higgsino ($H_{1/2}$) scenario}
\label{subsec:higgsino_scenario}

The first scenario for a degenerate pair of EW new states, the scenario $H_{1/2}$,
is provided by the Higgsino sector with the spin-1/2 SUSY partners
of the down- and up-type Higgs bosons in the MSSM. This is realized practically when
the Higgsino mass parameter $\mu$ of the superpotential term
$\mu \hat{H}_d\cdot \hat{H}_u$ mixing the two Higgs superfields is much smaller
than all the other SUSY parameters including the gaugino mass parameters,
$M_{1,2,3}$ \cite{Chen:1995yu,Giudice:2010wb,Han:2013kza,Cheung:2005pv,Baer:2011ec,
Baer:2012uy,Berggren:2013vfa}. (Without any loss of generality, we assume
the parameters, $M_{1,2}$ and $\mu$ to be real and positive in the present note.)
When the gaugino states as well as the other SUSY states are decoupled without
generating any mixing with the Higgsinos, the two SU(2)-doublet
Higgsino states
$\tilde{H}_d=[\tilde{H}^0_{dL}, \tilde{H}^-_{dL}]$  and
$\tilde{H}_u=[\tilde{H}^+_{uL},\tilde{H}^0_{uL}]$  have maximal mixing.
The mass term for the charged and neutral Higgsino states can be
cast into the mass term for a degenerate pair of a Dirac chargino and
a Dirac neutralino with mass $\mu$ as
\begin{eqnarray}
  \mu\left(\overline{\tilde{H}^{-}_{uR}} \tilde{H}^-_{dL}
          +\overline{\tilde{H}^+_{dR}} \tilde{H}^{+}_{uL} \right)
 -\mu \left(\overline{\tilde{H}^0_{uR}} \tilde{H}^0_{dL}
          +\overline{\tilde{H}^0_{dR}} \tilde{H}^0_{uL} \right)\quad
\Rightarrow \quad
  \mu\, \overline{\chi^-_H}\, \chi^-_H
 +\mu\, \overline{\chi^0_H}\, \chi^0_H
\end{eqnarray}
where the Dirac chargino and Dirac neutralino are defined by
\begin{eqnarray}
\chi^-_H \,= \, \tilde{H}^-_{dL} + \tilde{H}^-_{uR} \quad \mbox{and}\quad
\chi^0_H \,= \, \tilde{H}^0_{dL} -\tilde{H}^0_{uR}
\end{eqnarray}
in terms of the current Higgsino states with the charge-conjugated
states, $\tilde{H}^-_{uR}=(\tilde{H}^+_{uL})^c$ and
$\tilde{H}^0_{uR}=(\tilde{H}^0_{uL})^c$.

As the down- and up-type Higgsinos form a vector-like
SU(2)$_L$ doublet, the interactions of the Dirac chargino $\chi^-_H$ and Dirac
neutralino $\chi^0_H$ with the electromagnetic (EM) and weak gauge bosons are
described by the Lagrangian
\begin{eqnarray}
  {\cal L}^H_{V\chi\chi}
=  e\, \overline{\chi^-_H}\gamma^\mu \chi^-_H\, A_\mu
   + e\, \frac{(1/2-s^2_W)}{c_W s_W}\, \overline{\chi^-_H}\gamma^\mu \chi^-_H\, Z_\mu
   - \onehalf\, \frac{e}{c_W s_W}\, \overline{\chi^0_H}\gamma^\mu \chi^0_H\, Z_\mu
  -\frac{e}{\sqrt{2} s_W} \left( \overline{\chi^0_H}\gamma^\mu \chi^-_H\, W^+_\mu
                                  +{\rm h.c.}
                                  \right)
\end{eqnarray}
where the Lorentz structure of every gauge interaction term is of a pure vector type
and its strength is fixed only by the  positron electric charge $e$ and weak
mixing angle $\theta_W$. In the present note we use the abbreviations
$s_W=\sin\theta_W$ and $c_W =\cos\theta_W$ for the sake of convenience.

\subsection{The spin-1/2 wino ($W_{1/2}$) scenario}
\label{subsec:wino_scenario}

The second scenario for a degenerate pair of SUSY states, the $W_{1/2}$
scenario, is provided by the MSSM wino sector with the spin-1/2 partners of
the SU(2)$_L$ gauge bosons. This is realized practically when the SU(2)$_L$
gaugino mass parameter $M_2$ is much smaller than the other gaugino mass
parameters $M_{1,3}$ and the Higgsino mass parameter $\mu$
as well as all the other SUSY
parameters \cite{Giudice:2010wb,Han:2013kza,Cheng:1998hc,
Gherghetta:1999sw,Cheung:2005ba,Feng:1999fu,Ibe:2006de}. In this scenario the
mass term of the SU(2)-triplet wino state $\tilde{W}=[\tilde{W}^+_L,\tilde{W}^0_L,
\tilde{W}^-_L]$ can be cast into a Dirac mass term for a Dirac chargino and
a Majorana mass term for a Majorana neutralino with a common mass $M_2$ as
\begin{eqnarray}
M_2\, (\overline{\tilde{W}^+_R}\tilde{W}^+_L
    +\overline{\tilde{W}^0_R}\tilde{W}^0_L
    +\overline{\tilde{W}^-_R}\tilde{W}^-_L)
    \quad \Rightarrow \quad
     M_2\,\overline{\chi^-_W}\,\chi^-_W
    + \onehalf\, M_2\, \overline{\chi^0_W}\,\chi^0_W
\end{eqnarray}
by defining a Dirac chargino $\chi^-_W$ and a Majorana neutralino $\chi^0_W$ by
\begin{eqnarray}
\chi^-_W = \tilde{W}^-_L + \tilde{W}^-_R \quad \mbox{and}\quad
\chi^0_W = \tilde{W}^0_L + \tilde{W}^0_R
\end{eqnarray}
with the charge-conjugated states $\tilde{W}^\pm_R =(\tilde{W}^\mp_L)^c$ and
$\tilde{W}^0_R = (\tilde{W}^0_L)^c$. Note that by definition the neutralino state
is identical to its charge-conjugated anti-particle, i.e. $({\chi}^{0}_W)^c
= \chi^0_W$.

In the $W_{1/2}$ scenario, the interactions of the vector-like SU(2)-triplet
states with the EM and weak gauge bosons are described by
\begin{eqnarray}
  {\cal L}^W_{V\chi\chi}
=  e\, \overline{\chi^-_W}\gamma^\mu \chi^-_W\, A_\mu
   + e\, \frac{(1-s^2_W)}{c_W s_W}\, \overline{\chi^-_W}\gamma^\mu \chi^-_W\, Z_\mu
  -\frac{e}{s_W} \left( \overline{\chi^0_W}\gamma^\mu \chi^-_W\, W^+_\mu
                                  +{\rm h.c.}
                                  \right)
\end{eqnarray}
Again, like the $H_{1/2}$ scenario, the Lorentz structure of every gauge
interaction term is of a pure vector type, but the coupling strengths determined
uniquely by the weak mixing angle $\theta_W$ are characteristically different
from those in the $H_{1/2}$ scenario. Note that the Majorana neutralino in the
$W_{1/2}$ scenario couples neither to the photon nor to the neutral weak boson $Z$.

\subsection{The left-handed slepton ($L_0$) scenario}
\label{subsec:slepton_scenario}

The third scenario for a degenerate pair of SUSY states, the  $L_0$ scenario,
is provided by the MSSM left-handed slepton sector with the spin-0 partners
$\tilde{L}=[\tilde{\nu}_\ell, \tilde{\ell}^-_L]$ of the
SU(2)$_L$-doublet lepton.
This is realized practically when the SUSY-breaking slepton
mass parameter $\tilde{m}_{\ell_L}$ is much smaller than all the other SUSY parameters.
In general, the charged slepton $\tilde{\ell}^-_L$ and the sneutrino $\tilde{\nu}_\ell$
are non-degenerate and split by the so-called $D$-term potential after EWSB
$\Delta m^2 = m^2_{\tilde{\ell}^-_L}-m^2_{\tilde{\nu}_\ell}
=- m^2_Z \cos 2 \beta\, c^2_W$, vanishing for $\tan\beta=1$. For the sake of
comparison, the charged slepton and neutral sneutrino may be assumed to be
degenerate with $\tan\beta=1$ at the tree level.

In the $L_0$ scenario, the interactions of the left-handed SU(2)-doublet slepton
state with the EM and weak gauge bosons are described by the Lagrangian
\begin{eqnarray}
  {\cal L}^L_{V\tilde{\ell}_L\tilde{\ell}_L}
=  e\, \tilde{\ell}_L^+\overleftrightarrow{\partial_\mu} \tilde{\ell}_L^-\, A^\mu
   + e\, \frac{(1/2-s^2_W)}{c_W s_W}\,
     \tilde{\ell}_L^+\overleftrightarrow{\partial_\mu} \tilde{\ell}_L^-\, Z^\mu
   - \onehalf\, \frac{e}{c_W s_W}\,
      \tilde{\nu}_\ell^*\overleftrightarrow{\partial_\mu} \tilde{\nu}_\ell \, Z^\mu
  -\frac{e}{\sqrt{2} s_W}
    \left( \tilde{\nu}_\ell^*\overleftrightarrow{\partial_\mu} \tilde{\ell}_L^- \,
           W^{+\mu} +{\rm h.c.} \right)
\end{eqnarray}
where $A\overleftrightarrow{\partial_\mu}B=A \partial_\mu B - (\partial_\mu A) B$.
Note that the gauge coupling strengths of the charged slepton and neutral sneutrino
are identical to those of the Dirac chargino and Dirac neutralino in the
Higgsino case.
However, because of their zero spin values, the Lorentz structure of the gauge
interactions are different from that of the chargino and neutralino states.
In addition, there exist 4-point contact gauge interactions of left-handed sleptons.
The Lagrangian for the $\gamma \gamma \tilde{\ell}^-_L\tilde{\ell}^-_L$ and
$\gamma Z \tilde{\ell}^-_L\tilde{\ell}^-_L$ four-point vertices read
\begin{eqnarray}
  {\cal L}^L_{\gamma Z \tilde{\ell}^-_L\tilde{\ell}^-_L}
=  e^2 \tilde{\ell}^+_L \tilde{\ell}^-_L A_\mu A^\mu
 + 2 e^2 \frac{(1/2-s^2_W)}{c_W s_W} \tilde{\ell}^+_L \tilde{\ell}^-_L A_\mu Z^\mu
\end{eqnarray}
Because of these momentum-independent contact terms the charged slepton pair production
associated with a hard final-state as well as initial-state photon emission exhibits
a $S$-wave threshold excitation pattern in contrast to the $P$-wave excitation pattern
in the neutral sneutrino pair production only with a hard initial photon emission,
as shown later in Sec.$\,$\ref{subsec:fsr}.

\subsection{Feynman rules for a vector boson converting into a particle pair
            and a photon}
\label{subsec:feynman_rules}

Depending on the electric charge and spin of the SUSY EW particle $X$,
the vertex
$VX\bar{X}$ for the process $V^*(q)\to X(q_1) \bar{X}(q_2)$ with $V=\gamma, Z$
can be parameterized as
\begin{eqnarray}
  \langle X(q_1)\bar{X}(q_2)\, ||\, V^\mu(q)\rangle
\, = \, e\, c^V_X \, \left\{\begin{array}{cl}
   (q_1-q_2)^\mu        &  \quad \mbox{for spin-0 chaged sleptons or sneutrinos} \\[2mm]
   \bar{u}(q_1)\, \gamma^\mu\, v(q_2)
                        &  \quad \mbox{for spin-1/2 charginos or neutralinos}
             \end{array}\right.
\label{eq:vxx_vertex}
\end{eqnarray}
with $q=q_1+q_2$ and the normalized
couplings $c^V_X$ for $(V=\gamma, Z)$ expressed as
\begin{eqnarray}
&& c^\gamma_{\chi^-_H} = c^\gamma_{\tilde{\ell}^-_{L}} =  1, \quad
   c^Z_{\chi^-_H} = c^Z_{\tilde{\ell}^-_L} = \frac{(1/2-s^2_W)}{c_W s_W}, \quad
   c^Z_{\chi^0_H} = c^Z_{\tilde{\nu}_\ell} = -\frac{1}{2 c_W s_W}
   \label{eq:three_point_couplings_1} \\
&& c^\gamma_{\chi^-_W} = 1, \quad c^Z_{\chi^-_W} = \frac{c_W}{s_W}, \quad
   c^Z_{\chi^0_W} = 0
  \label{eq:three_point_couplings_2}
\end{eqnarray}
in terms of $c_W$ and $s_W$.

In addition to the standard three-point vertices in
Eqs.$\,$(\ref{eq:three_point_couplings_1}) and (\ref{eq:three_point_couplings_2}),
there exists a four-point momentum-independent vertex contributing to the FSR
process $V^* \to \gamma\, \tilde{\ell}^-_L \tilde{\ell}^+_L $ in the $L_0$ scenario:
\begin{eqnarray}
\langle \gamma^\nu  \tilde{\ell}^+_L\tilde{\ell}^-_L\, ||\, V^\mu \rangle
 = 2 e^2\, d^{V}_{\tilde{\ell}^-_L} g^{\mu\nu}
\label{eq:vgll_vertex}
\end{eqnarray}
with the normalized couplings $d^{\gamma,Z}_{\tilde{\ell}^-_L}$ identical to
$c^{\gamma,Z}_{\tilde{\ell}^-_L}$ given in
Eq.$\,$(\ref{eq:three_point_couplings_1}).

\subsection{Radiatively-induced mass difference}
\label{subsec:mass_difference}

Although in all the three scenarios the charged and neutral SUSY particles are
degenerate in mass before EWSB, the gauge symmetry breaking part in the MSSM causes
a finite calculable mass splitting through radiative corrections. Moreover,
the so-called $D$-term potential leads to an additional mass
splitting
between the spin-0 charged slepton and neutral sneutrino in the $L_0$ scenario
unless the two Higgs vacuum expectation values, $v_u=v \cos\beta$ and
$v_d=v \sin\beta$, are equal, i.e. $\tan\beta=1$.

At the leading order the mass splitting stems from one-loop virtual photon and
$Z$-boson exchange corrections  to the masses and the wave functions of the
chargino and neutralino states in the $H_{1/2}$ or $W_{1/2}$
scenarios \cite{Cheung:2005pv,Gherghetta:1999sw,Cheung:2005ba,Feng:1999fu,
Ibe:2006de,Dimopoulos:1990kc,Thomas:1998wy,Sher:1995tc,Pierce:1996zz}.
The one-loop mass splitting for the on-shell SUSY states is
\begin{eqnarray}
\Delta m_H &=& m_{\chi^\pm_H}-m_{\chi^0_H}
          = \frac{\alpha}{4\pi} \mu \left[ f(m_Z/\mu)-f(0) \right] \\
\Delta m_W &=& m_{\chi^\pm_W}-m_{\chi^0_W}
          = \frac{\alpha}{4\pi s^2_W} M_2
            \left[ f(m_W/M_2)-c^2_W f(m_Z/M_2)- s^2_W f(0)\right]
\end{eqnarray}
respectively, where the loop function $f(a)=2 \int^1_0 dx\, (1+x) \ln \left[x^2+(1-x)
a^2\right]$ and $\alpha=e^2/4\pi$. The asymptotic value of the mass splitting for $\mu, M_2 \gg m_Z$ is
$\alpha m_Z/2 \simeq 355$ MeV and $\alpha m_W/2(1+c_W)\simeq 165$ MeV, respectively.

In the $L_0$ scenario, the charged slepton is in general non-degenerate with the
neutral sneutrino, the SU(2)$_L$ doublet partner, due to the $D$-term contribution
leading to a mass splitting of $O(m^2_Z/M_s)$ where $M_s$ is a common SUSY-breaking
slepton mass parameter, unless $\tan\beta=1$. Even if they are degenerate with
$\tan\beta=1$ at the tree level, a leading-order mass splitting arises from one-loop
corrections with virtual sleptons of same and different flavor and Higgs bosons
as well as virtual photon, $Z$-boson and $W$-boson diagrams. Nevertheless, as the
splitting must vanish without EWSB, it is therefore bounded by a quantity proportional
to the EM fine structure constant times the $Z$-boson mass.


\section{Single-photon processes at $e^+e^-$ colliders}
\label{sec:single-photon_processes}

In the above scenarios, the pair of SUSY states may be produced at the ILC
via $s$-channel $\gamma/Z$ exchanges. However, as the mass splitting between the
charged and neutral states is of the order of a few hundred MeV,
the expected signatures at the ILC can vary from soft ($p_T \sim 300~\mathrm{MeV}$)
decay products through displaced vertices to massive charged tracks. We do not
perform any sophisticated analyses for distinguishing the charged modes from
the neutral modes in the present work and assume the charged and neutral states
in each scenario to be (nearly) degenerate in the following numerical
analyses.
For more dedicated studies for separating the charged modes from the neutral modes
and measuring the mass splitting based on the visible decay products of the charged
states, we refer to Refs.$\,$\cite{Hensel:2002bu,Berggren:2013vfa}.

\begin{figure}[tb]
\centering
\epsfig{file=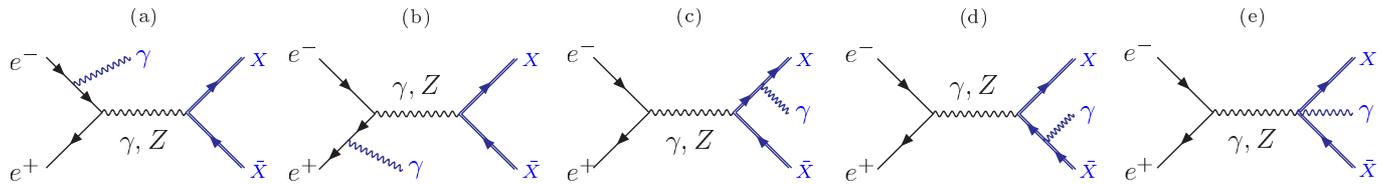,width=\textwidth}
\caption{Feynman diagrams for the single-photon process $e^+e^-\to\gamma X\bar{X}$
    with the charged or neutral particle-antiparticle pair, $X$ and $\bar{X}$.
    The diagrams (a) and (b) are for the ISR processes with the photons radiated
    from the initial electron and positron lines with $X=\{\chi^-_{H,W}, \chi^0_H,
    \tilde{\ell}^-_L,\tilde{\nu}_\ell\}$ and the diagrams (c) and (d) for the FSR
    processes with the photons emitted from the final-state charged particles with
    $X=\{\chi^-_{H,W}, \tilde{\ell}^-_L\}$. The diagram (e) involving a four-point coupling
    is only for a scalar particle $X=\tilde{\ell}^-_L$.
    }
\label{fig:Feynman_diagrams}
\end{figure}

One method to search for production of invisible particles is to identify an
associated hard radiated photon in single-photon processes in $e^+e^-$ collisions,
$e^+e^-\to\gamma +\slashed{E}$.
In the three $\{H_{1/2}, W_{1/2}, L_0\}$ scenarios, a pair of charged or neutral
particles, $X\bar{X}$, are produced through a virtual $\gamma$ or $Z$-exchange
and accompanied by a hard photon radiation in the single-photon process
\begin{eqnarray}
e^+  e^-\,\to\, \gamma\ V^* \ \ \mbox{or} \ \ V^*
        \, \to\, \gamma \ X  \bar{X} \quad
\mbox{with}\quad V=\gamma, Z
\label{eq:sigle_photon_process}
\end{eqnarray}
For the neutral $\chi^0_H\chi^0_H$ and $\tilde{\nu}_\ell \tilde{\nu}^*_\ell$ pairs,
the photon in the single-photon process (\ref{eq:sigle_photon_process})
is radiated only from the initial electron or positron line, but for every
charged pair the photon is emitted also from the final charged particle lines
as shown in Fig.$\,$\ref{fig:Feynman_diagrams}. In each process, the ISR
and FSR parts are separately EM gauge invariant and develop no interference
terms between them (when the $Z$-boson width is ignored).

The FSR part has been ignored in most of the previous studies on the single-photon
processes. In the present analysis we include not only the ISR part but also
the FSR part for assessing the validity of the ISR approximation and the influence
of the FSR part in characterizing the (nearly) invisible particles through
single-photon processes in $e^+e^-$ collisions.

\subsection{Initial state radiation}
\label{subsec:isr}

We ignore the electron mass
except for avoiding collinear singularity. We include the possible $e^\pm$ beam polarizations $P_\pm$ in studying the dependence of the signal process $e^+e^-\to \gamma X\bar{X}$ on the photon energy
fraction $x_\gamma=2E_\gamma/\sqrt{s}$ and the photon scattering angle
$\theta_\gamma$ with respect to the $e^-$ momentum direction in the $e^+e^-$
center-of-mass (CM) frame.

The ISR effect can be expressed in a factorized form with a universal ISR radiator
function \cite{Nicrosini:1989pn,Montagna:1995wp,Berends:1984dw} as
\begin{eqnarray}
  \frac{d\sigma  (e^+e^-\to\gamma X\bar{X})_{\rm ISR}}{dx_\gamma\, d\cos\theta_\gamma}
\, =\, {\cal R}(s; x_\gamma, \cos\theta_\gamma)
       \times \sigma^{X\bar{X}} (q^2)
\label{eq:isr_cross_section}
\end{eqnarray}
where the ISR radiator function ${\cal R}$ can be expressed to a very good
approximation as
\begin{eqnarray}
  {\cal R}(s;x_\gamma, \cos\theta_\gamma)
= \frac{\alpha}{\pi}\frac{1}{x_\gamma}
  \left[\frac{1+(1-x_\gamma)^2}{1+4 m^2_e/s -\cos^2\theta_\gamma}
       -\frac{x^2_\gamma}{2}\right]
\end{eqnarray}
which is nearly independent of the beam energy except for the forward or
backward collinear direction. The total cross section of the $X\bar{X}$ pair
production in $e^+e^-$ annihilation to be evaluated with the reduced CM energy
squared $q^2=(1-x_\gamma) s$ is given by
\begin{eqnarray}
  \sigma^{X\bar{X}}(q^2)
= \frac{2\pi\alpha^2}{3 {q^2}} \beta_{q} {\cal P}(X; P_-, P_+; q^2) \,
  {\cal K}(\beta_{q})
\end{eqnarray}
with $\beta_{q}=\sqrt{1-4 m^2_X/(1-x_\gamma)s}$, the speed
of the particle $X$ in the $X\bar{X}$ CM frame. The polarization-dependent
factor ${\cal P}$ is defined in terms of the beam polarizations and $\gamma$ and
$Z$-boson propagators as
\begin{eqnarray}
 {\cal P}(X; P_-,P_+; q^2)
=  \frac{(1+P_-)(1-P_+)}{4}
     \left|\, c^\gamma_{X}+c_R c^Z_{X} \frac{q^2}{q^2- m^2_Z}\right|^2
 + \frac{(1-P_-)(1+P_+)}{4}
     \left|\, c^\gamma_{X}+c_L c^Z_{X} \frac{q^2}{q^2- m^2_Z}\right|^2
\label{eq:polarization_factor}
\end{eqnarray}
with $c_L=(1/2-s^2_W)/c_Ws_W$ and $c_R=-s_W/c_W$ and the couplings $c^{\gamma}_X$
and $c^Z_X$ given in Eqs.$\,$(\ref{eq:three_point_couplings_1}) and
(\ref{eq:three_point_couplings_2}). The kinematical factor
${\cal K}(\beta_q)$ reads
\begin{eqnarray}
  {\cal K}({\beta_q })
= \left\{\begin{array}{cl}
         \beta^2_{q}       & \qquad
         \mbox{{for} spin-0 charged slepton or sneutrino} \\[2mm]
         2 (3-\beta^2_{q}) & \qquad
         \mbox{{for} spin-1/2 chargino or neutralino}
         \end{array}\right.
\label{eq:kinematical_factor}
\end{eqnarray}
The range of $x_\gamma$ is $0\leq x_\gamma \leq 1- 4m^2_X/s$ with its maximal
value $x^{max}_\gamma=1- 4m^2_X/s$ corresponding to the $X\bar{X}$
production threshold with $\beta_{q}=0$. Asymptotically when
$\beta_{q}\to 0$, i.e. $x_\gamma\to 1 - 4 m^2_X/s$, the cross section
is proportional to $\beta^3_{q}$ for the spin-0 particles, exhibiting a
slowly-rising $P$-wave threshold excitation, but it is proportional to
$\beta_{q}$ for the spin-1/2 particles, exhibiting a steeply-rising
$S$-wave excitation near the threshold.

\subsection{Final state radiation}
\label{subsec:fsr}

Unlike the ISR effect, the FSR parts of the photon-energy and angular distributions
are not universal and have no collinear singular term.

For any charged pair $X\bar{X}=\chi^-_H\chi^+_H, \chi^-_W\chi^+_W,
\tilde{\ell}^-_L\tilde{\ell}^+_L$, the dependence of the FSR part on the FSR
photon energy fraction $x_\gamma$ and the photon scattering angle $\theta_\gamma$
can be decomposed as
\begin{eqnarray}
  \frac{d\sigma  (e^+e^-\to\gamma X\bar{X})_{\rm FSR}}{dx_\gamma\, d\cos\theta_\gamma}
\, = \,\frac{3}{8}\, \left[ (1+\cos^2\theta_\gamma) {\cal F}^{X}_1(s; x_\gamma)
              + (1-3\cos^2\theta_\gamma) {\cal F}^{X}_2(s; x_\gamma) \right]
              \times \sigma^{X\bar{X}} (s)
\label{eq:fsr_cross_section}
\end{eqnarray}
where the final-state radiator functions ${\cal F}^X_{1,2}$ are process-dependent.
Explicitly, for the production of a chargino pair with $X
=\chi^-_H$ or $\chi^-_W$, the FSR radiator functions are given by
\begin{eqnarray}
    {\cal F}^X_1 (s;x_\gamma)
&=& \frac{\alpha}{\pi}\frac{1}{x_\gamma}\frac{{\beta_q }}{{\beta_s }}
    \left[(1+{\beta_s^2 } - 2 x_\gamma) L({\beta_q })
          - 2 (1-x_\gamma) +\frac{2x^2_\gamma}{3- {\beta_s^2 }}
           [L({\beta_q })-1]\right] \\
    {\cal F}^X_2 (s;x_\gamma)
&=& \frac{\alpha}{\pi}\frac{1}{x_\gamma}\frac{{\beta_q }}{{\beta_s }}
    \frac{2}{3-{\beta_s^2 }}
    \left[2-2x_\gamma - (1-{\beta_s^2 }) L({\beta_q })\right]
\end{eqnarray}
in terms of $x_\gamma$ with ${\beta_s } =\sqrt{1-4 m^2_X/s}$, the CM speed
of the $X$ in the process $e^+e^-\to X\bar{X}$ with no photon
emission \cite{Dokshitzer:1994jt}.
On the other hand, for the production of a charged slepton pair with
$X=\tilde{\ell}^-_L$, the FSR radiator functions read
\begin{eqnarray}
    {\cal F}^X_1 (s;x_\gamma)
&=& \frac{\alpha}{\pi}\frac{1}{x_\gamma}\frac{{\beta_q }}{{\beta_s }}
    \left[(1+{\beta_s^2 }-2 x_\gamma) L({\beta_q }) - 2 (1-x_\gamma)
          + \frac{2x^2_\gamma}{{\beta_s^2 }}\right] \\
    {\cal F}^X_2 (s;x_\gamma)
&=& \frac{\alpha}{\pi}\frac{1}{x_\gamma}\frac{{\beta_q }}{{\beta_s }}
    \frac{1}{{\beta_s^2 }}
    \left[(3-{\beta_s^2 }-2x_\gamma) L({\beta_q })
    - 6(1-x_\gamma)\right]
\end{eqnarray}
with the logarithmic function $L({\beta_q })$ defined by
\begin{eqnarray}
  L({\beta_q })
= \frac{1}{{\beta_q }}
  \ln\left(\frac{1+{\beta_q }}{1-{\beta_q }}\right)
\end{eqnarray}
Integrating the distribution over the full range of the photon scattering angle,
the normalized FSR-photon energy distribution approaches a well-known universal
FSR radiator function in the soft-photon limit with $x_\gamma$ close to zero:
\begin{eqnarray}
  {\cal F}^{X}_1(s; x_\gamma)\, \rightarrow \,
  \frac{\alpha}{\pi}\frac{1}{x_\gamma}
  \left[(1+{\beta_s^2 }) L({\beta_s }) - 2\right] \quad
  \mbox{as}\quad x_\gamma\to 0
\end{eqnarray}
independently of the spin of the charged particle emitting the
photon \cite{Low:1958sn,Burnett:1967km}.

When the photon energy fraction approaches the $X\bar{X}$
threshold, the radiator function ${\cal F}^X_2$ goes to zero $\sim \beta^3_{q}$
for both the spin-0 and spin-1/2 cases. In contrast to this $P$-wave behavior,
the radiator function ${\cal F}^X_1$ exhibits a $S$-wave threshold behavior
as
\begin{eqnarray}
  {\cal F}^X_1(s;x_\gamma)\ \ \rightarrow \ \
  \frac{\alpha}{\pi}\, {\beta_q }
  \left\{\begin{array}{ll}
         2/\beta_s              & \quad  \mbox{for spin-0 charged slepton} \\[3mm]
         2\beta_s/(3-\beta_s^2) & \quad
         \mbox{for spin-1/2 chargino}
         \end{array} \right.
\qquad \mbox{as} \ \ x_\gamma \to {\beta_s^2 }
\end{eqnarray}
not only for the spin-1/2 chargino case but also for the spin-0 charged slepton case.
In the charged slepton case, the $S$-wave excitation of the FSR part is due to the
momentum-independent four-point contact terms contributing to the diagram
in Fig.$\,$\ref{fig:Feynman_diagrams}(e).

\subsection{Effects of the ISR and FSR in charged pair production}
\label{subsec:isr_fsr_comparison}

The FSR part in the photon-associated charged pair production is expected to be
much smaller in magnitude than the ISR part as the photon in the FSR part is
generated from a charged particle much heavier than the electron. Because of this
generally-expected feature, the FSR part has been ignored in most previous
analytic and numerical analyses on the single-photon processes. In this subsection,
we assess the validity of the ISR approximation critically by exploiting
the ratio of the FSR part to the ISR part defined as
\begin{eqnarray} \label{eq:FI}
  {\cal R}_{\rm FI} (x_\gamma)
= \frac{d\sigma(e^+ e^- \to \gamma X\bar{X})_{\rm FSR}/dx_\gamma}{
        d\sigma(e^+ e^- \to \gamma X\bar{X})_{\rm ISR}/dx_\gamma}.
\end{eqnarray}
in terms of the $x_\gamma$-dependent distributions derived
by integrating Eqs.(\ref{eq:isr_cross_section})
and (\ref{eq:fsr_cross_section}) over the scattering angle $\theta_\gamma$,
respectively.

\begin{figure}[tb]
\epsfig{file=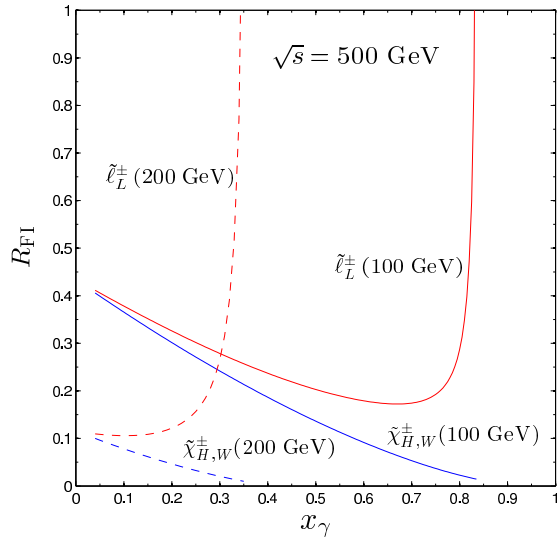,width=0.4\textwidth}
\caption{Ratio of the FSR to the ISR versus $x_\gamma$ in the production of a
         charged pair at a 500 GeV ILC. The solid and  dashed lines
         are for $m_X=100$ GeV and 200 GeV, respectively. The up-ward
         (red) and falling (blue) lines are for the spin-0 charged slepton case
         and the spin-1/2 chargino cases, respectively.
         }
\label{fig:ratio_fsr_to_isr}
\end{figure}

Figure~\ref{fig:ratio_fsr_to_isr} shows the dependence of the ratio
of the FSR part to the ISR part for two mass values, $m_X=100$ GeV (solid lines) and
200 GeV (dashed lines). The photon scattering angle has been restricted to
$10^\circ < \theta_\gamma <170^\circ$.  As the falling (blue) lines indicate,
the FSR part of the chargino
pair production cross section is consistently smaller than the corresponding
ISR part and it becomes negligible, in particular, near the threshold.
As the mass increases, the ratio is even more suppressed.
Nevertheless, for more precise mass and coupling measurements it will be more
meaningful to include the FSR part in any realistic analyses.

In contrast to the spin-1/2 chargino case, the ratio of the FSR part to the ISR part
does not monotonically decrease with increasing $x_\gamma$ in the slepton scenario.
In fact, the ratio blows up near the threshold, as the FSR part decreases in proportion
to $\beta_q $ in $S$ waves while the ISR part decreases in proportion
to $\beta_q^3$ in $P$ waves. Therefore, the FSR part needs to be
definitely included for any studies based on the threshold behavior of the
production cross section.

\section{Characteristics of the production cross sections}
\label{sec:characteristics}

\begin{figure}[tb]
\centering
\epsfig{file=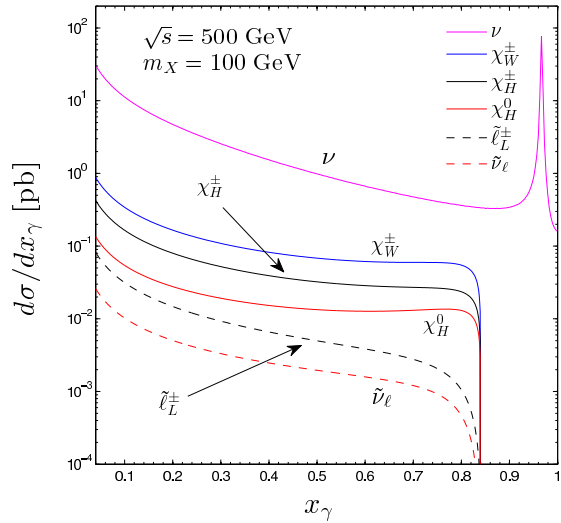,width=0.4\textwidth}
\caption{Unpolarized $x_\gamma$ distribution $d\sigma/d x_\gamma$
         with $m_X = 100$ GeV at a 500 GeV ILC, for different SUSY EW
         particles, as well as that of the background process $e^+e^-\to
         \gamma\nu\bar{\nu}$ (solid line on the top).
         The photon scattering angle has been restricted to
         $10^\circ < \theta_\gamma <170^\circ$. }
\label{fig:extra1}
\end{figure}

The most severe irreducible background to the signal events under consideration is
the standard $e^+e^-\to \gamma \nu\bar{\nu}$ with $\nu=\nu_e, \nu_\mu$ and $\nu_\tau$. For the sake of comparison, the unpolarized $x_\gamma$ distribution
for the background is shown (solid line on the top) together with the
distributions for different SUSY EW particles with $m_X=100$ GeV in
Fig.$\,$\ref{fig:extra1}. Throughout this paper, we will illustrate our results
for a 500 GeV ILC.

For $m_X > m_Z/2$, one powerful kinematic cut for reducing the
irreducible background reaction $e^+e^-\to\gamma\nu\bar{\nu}$ can be applied to the
recoil mass squared $q^2 = (q_1+q_2)^2 = (p_1+p_2-k)^2=s(1-x_\gamma)$ which
can be very accurately reconstructed by measuring the photon energy fraction
$x_\gamma$. We evaluate the overall statistical significance $N_{SD}$ for the
signal and background by summing over all events not only with the photon energy and
angular cuts applied but also with the recoil mass cut $\sqrt{q^2} > 2 m_X$. Note
that this mass cut eliminates the $Z$-pole contribution to the $\gamma\nu\bar{\nu}$
background.

Another way of removing the background significantly is to exploit the electron and
positron beam polarizations. The $t$-channel $W$-exchange diagrams
contribute to the background process $e^+e^-\to \gamma \nu_e \bar{\nu}_e$ only for
the left-handed electrons so that the background can be significantly reduced by
taking the right-handed electron and left-handed positron beams. However, which
beam polarization is more efficient for the signal significance is determined
also according to the polarization dependence of the signal events.

\subsection{Statistical significance of signal events}
\label{subsec:significance}

In order to quantify whether an excess of signal photons from the $X\bar{X}$ pair
production, $N_S= {\cal L} \sigma $ for a given integrated luminosity ${\cal L}$,
can be measured over the $N_B = {\cal L} \sigma_B$ SM background photons from the
radiative neutrino production, we define a simple-minded theoretical significance
\begin{eqnarray}
N_{SD} \, =\, \frac{N_S}{\sqrt{N_S+N_B}}
  \, =\, \frac{\sigma}{\sqrt{\sigma+\sigma_B}}\, \sqrt{{\cal L}}
\end{eqnarray}
For our simple numerical analysis we require the photon energy to be
$E_\gamma > 10$ GeV, corresponding to $x_\gamma > 0.04$
and the photon scattering angle to be
$10^\circ < \theta_\gamma <170^\circ$ so as to guarantee that the photon will
have an accurate momentum measurement.  We also assume the CM energy $\sqrt{s}=500$ GeV and the total integrated luminosity ${\cal L}=500$ fb$^{-1}$.

\begin{figure}[tb]
\begin{center}
\epsfig{file=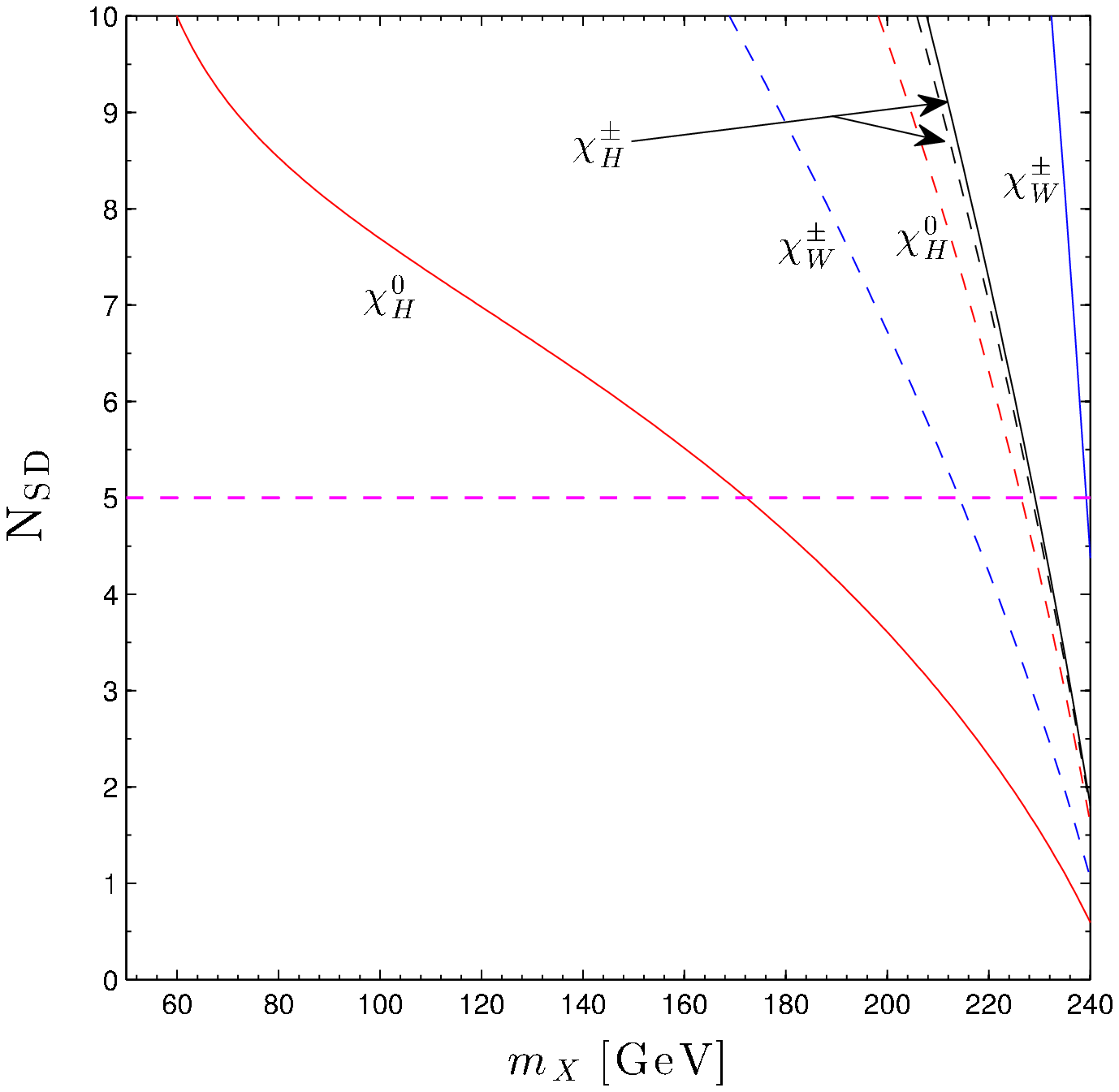,width=0.4\textwidth}\hskip 1 cm
\epsfig{file=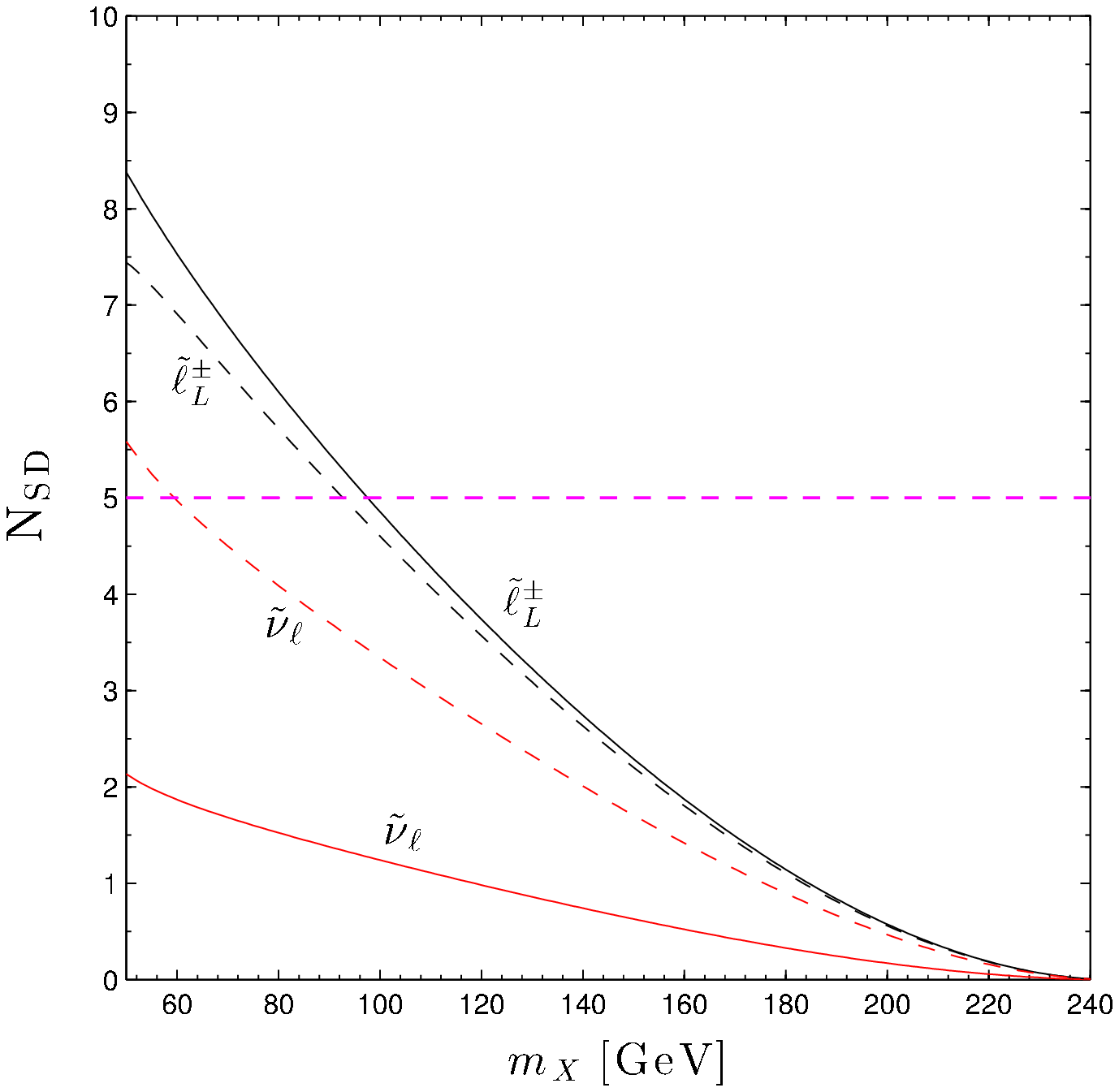,width=0.4\textwidth}
\end{center}
\caption{Statistical significance $N_{SD}$ versus $m_X$
         for $\sqrt{s}=500$ GeV and the total integrated luminosity
         ${\cal L}=500$ fb$^{-1}$. The left panel is for the spin-1/2 chargino or
         neutralino pair production and the right panel for the spin-0 slepton pair
         production. The solid/dashed lines are for the left-handed/right-handed
         electron and right-handed/left-handed positron beam polarizations with
         $(P_-, P_+)=(\mp 0.8, \pm0.3)$, respectively.
         }
\label{fig:significance}
\end{figure}

The number of signal events needed for
a required $N_{SD}$ depends not only on the beam polarization, but also on $m_X$,
since the recoil mass cut $\sqrt{q^2}=2m_X$ is
applied to the background process. For example, for $m_X=100$ GeV, the total cross section of the
background for $(P_-, P_+)= (-0.8, +0.3)$ is about 6230 fb  implying $N_S\sim$ 8840 signal events needed for  statistical significance $N_{SD}=5$, while for $(P_-, P_+)= (+0.8, -0.3)$ the cross section is 400 fb and only $N_S\sim$ 2250 signal events is enough to reach $N_{SD}=5$.

Figure \ref{fig:significance} shows the dependence of the signal significance
$N_{SD}$ on the mass $m_X$.
The left panel is for the spin-1/2 chargino or neutralino pair production and
the right panel is for the spin-0
slepton pair production. In each panel, the solid lines are for the left-handed
electron and right-handed positron beam polarizations with $(P_-, P_+)= (-0.8, +0.3)$
and the dashed lines for the right-handed electron and left-handed positron beam
polarization with $(P_-, P_+)= (+0.8, -0.3)$.

The value of the statistical significance $N_{SD}$ is very sensitive to the
beam polarizations in the wino-type chargino $\chi^\pm_W$ and Higgsino-type
neutralino $\chi^0_H$ cases.
As the red solid and dashed lines in the left panel indicate, the significance for
the Higgsino-type neutralino $\chi^0_H$ is enhanced with the
right-handed/left-handed
electron/positron beam polarizations. On the contrary, the significance for the
wino-type chargino $\chi^\pm_W$ is greatly enhanced with the left-hand/right-handed
electron/positron beam polarizations. In both $H_{1/2}$ and $W_{1/2}$ scenarios,
the neutralinos as well as charginos can be discovered with large statistical
significances up to their mass close to the beam energy $\sqrt{s}/2$.

In contrast, as shown in the right panel of Fig.$\,$\ref{fig:significance} the value
of the statistical significance for the charged slepton pair and the sneutrino
pair production is so small that the charged slepton and the neutral sneutrino
can be
discovered only when its mass is less than $\sim 100$~GeV and 60~GeV, respectively,
or the integrated luminosity is much larger.

\subsection{Spin determination}
\label{subsec:spin_determination}

\begin{figure}[tb]
\epsfig{file=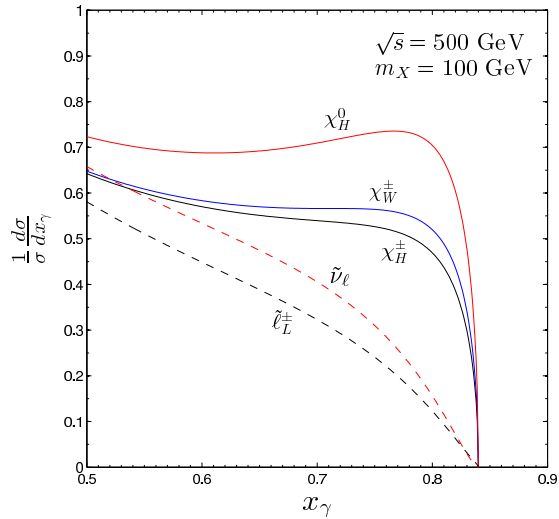,width=0.4\textwidth}
\caption{Normalized distribution versus the photon energy
         fraction $x_\gamma$ with $m_X=100$ GeV. Effects of both FSR and ISR are included.
         }
\label{fig:threshold_behavior}
\end{figure}

As indicated by the kinematical factor ${\cal K}(\beta_q)$ in
Eq.$\,$(\ref{eq:kinematical_factor}), the threshold behavior of the production
cross section of a neutral pair is distinctly different in the spin-0 and
spin-1/2 cases. As the red solid and dashed lines in
Fig.$\,$\ref{fig:threshold_behavior} show, the normalized cross section for
a spin-1/2 Higgsino-type Dirac neutralino pair is steeply excited in
$S$ waves at the threshold but the corresponding cross section for a spin-0 sneutrino
pair is slowly excited in $P$ waves. In this neutral pair production case,
the spin identification can be made unambiguously through the $x_\gamma$
distribution pattern near the threshold.

Like the neutral case, the ISR part of the production cross section for a charged
pair exhibits a $S$-wave and $P$-wave excitation for the spin-1/2 and spin-0
particle, respectively. As pointed out before, the FSR part is steeply excited
in $S$ waves even in the spin-0 case, which could spoil the characteristic spin-0
$P$-wave threshold behavior for the ISR part. However, as can be checked
quantitatively with the relative contribution of the FSR part in
Fig.$\,$\ref{fig:ratio_fsr_to_isr}, the FSR part becomes larger than the ISR
only when the photon energy fraction $x_\gamma$ is extremely close to the threshold
value, where both the FSR and ISR parts are already very small due to the
suppressed phase-space factor $\beta_q$. Therefore, as shown in
Fig.$\,$\ref{fig:threshold_behavior}, the spin of the SUSY EW particles can
be determined unambiguously through the excitation pattern of the (normalized) photon
energy distributions near the threshold - a sharp $S$-wave excitation for a
spin-1/2 particle and a slow $P$-wave excitation for a spin-0 particle,
only with a negligible contamination
of the FSR part even for the charged pair production.

\subsection{Ratio of left-handed and right-handed cross sections}
\label{subsec:ratios}

\begin{figure}[tb]
\centering
\epsfig{file=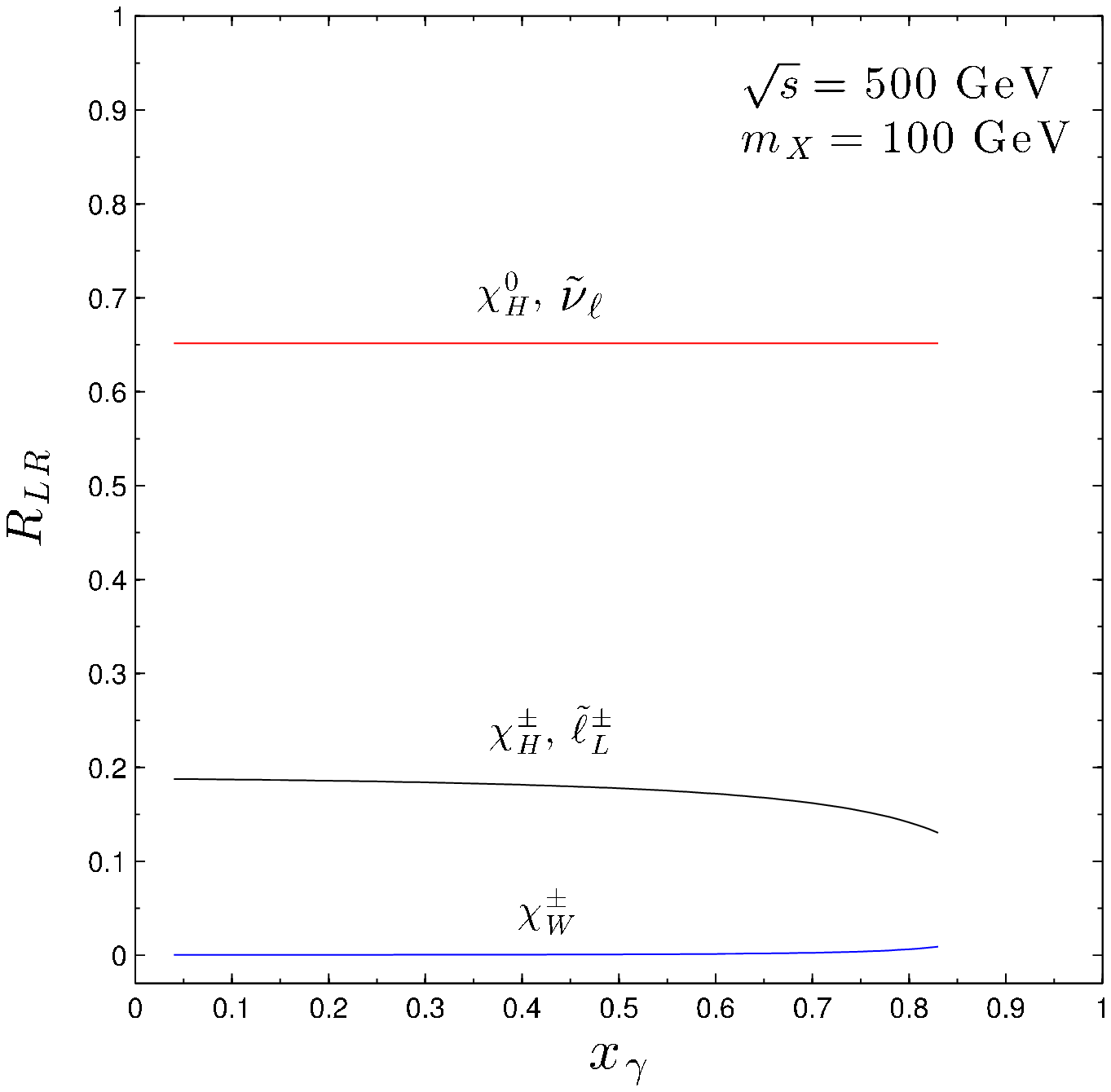,width=0.4\textwidth}\hskip 1.cm
\epsfig{file=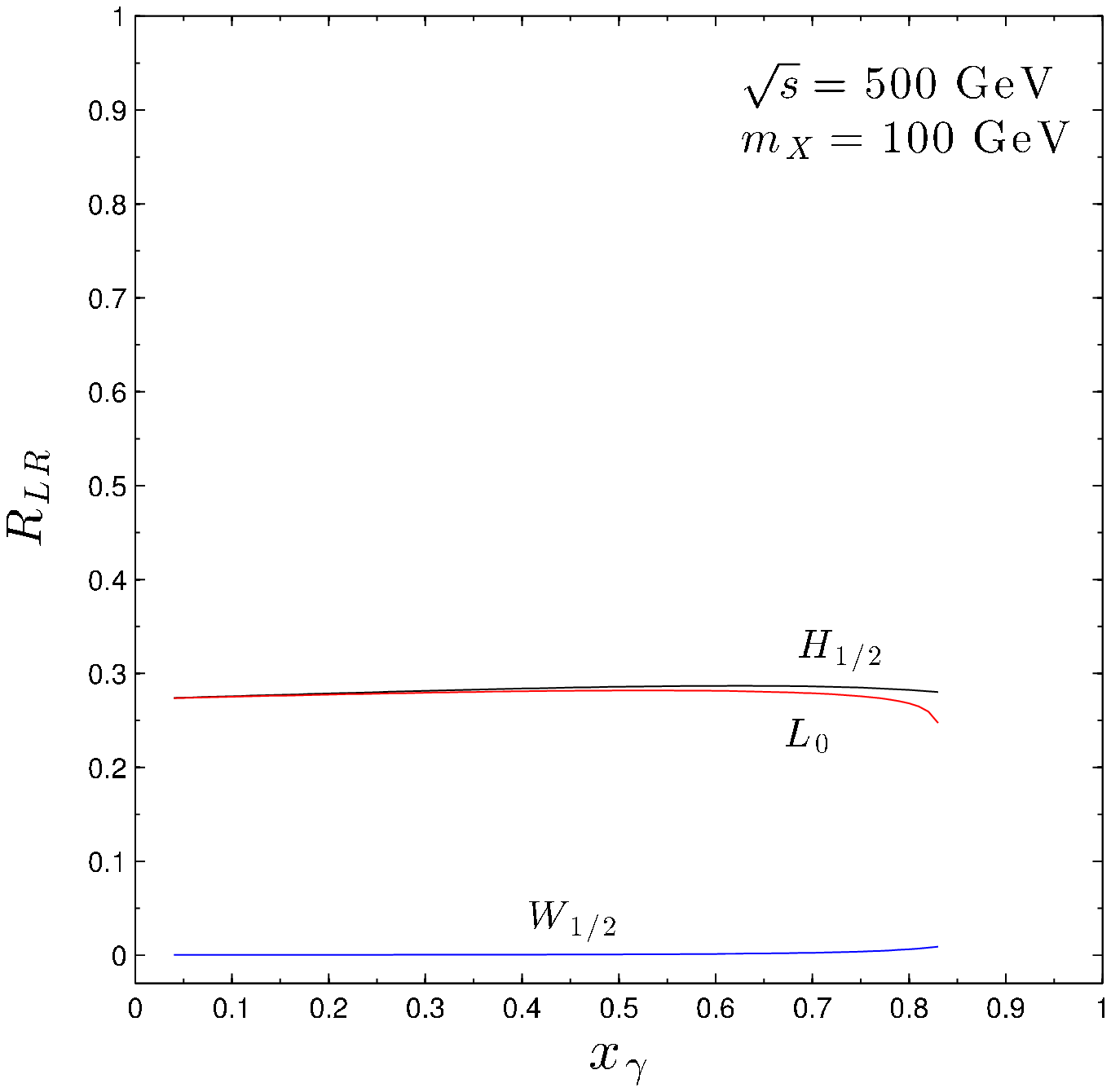,width=0.4\textwidth}
\caption{Ratio of the purely right-handed electron cross section
         to the purely left-handed electron cross section versus the photon energy
         fraction $x_\gamma$ with $m_X=100$ GeV. Left panel: individual channels of the pair production. Right panel:
          inclusive sums of the charged and neutral pair production.
          }
\label{fig:ratio_of_polarized_xsections}
\end{figure}

To see the polarization dependence of the signal cross sections, we define
the left-right (LR) ratio of the purely right-handed cross section to the
purely left-handed cross section:
\begin{eqnarray} \label{eq:LR}
  {\cal R}_{LR} (X;x_\gamma)
= \frac{d\sigma(e^+ e^-_R \to \gamma X\bar{X})/dx_\gamma}{
        d\sigma(e^+ e^-_L \to \gamma X\bar{X})/dx_\gamma}
\end{eqnarray}
obtained after applying the photon-angle cut described before.
Fig.$\,$\ref{fig:ratio_of_polarized_xsections} shows the $x_\gamma$ dependence of the
ratio of the right-handed electron cross section to the left-handed electron cross
section.

Before discussing the features that the LR ratios exhibit, we note that
for $m_X=100$ GeV the inequality relation $s\geq q^2 \geq 4 m^2_X
= 4\times 10^4\,\, {\rm GeV}^2 \gg m^2_Z$ is satisfied so that the polarization
factor ${\cal P}$ defined in
Eq.$\,$(\ref{eq:polarization_factor}) is nearly constant over the whole $x_\gamma$
range $[0.05, 0.84]$. In particular, for the neutral pair production with the photon
radiated from the initial electron or positron line and with no virtual-photon
exchange, the ratio is indeed constant and its value for the SU(2)$_L$ doublet state
$X=\chi^0_H, \tilde{\nu}_\ell$ is given by
\begin{eqnarray}
{\cal R}_{LR} [X] = \frac{c^2_R}{c^2_L} =\frac{s^4_W}{(1/2-s^2_W)^2} \simeq
 0.648
\quad\mbox{for}\quad X = \chi^0_D, \tilde{\nu}_\ell
\end{eqnarray}
independently of the spin of the produced particle $X$ for
$s^2_W\simeq 0.223$ given in Ref.$\,$\cite{Agashe:2014kda},
as shown in the left frame of Fig.$\,$\ref{fig:ratio_of_polarized_xsections}.

In contrast to the neutral pair production, the LR ratio for each charged pair
production exhibits a slight dependence on the photon energy fraction $x_\gamma$ with
a visible variation near the threshold with $x_\gamma=1-4m^2_X/s =0.84$ (see
the lower two lines in the left frame of
Fig.$\,$\ref{fig:ratio_of_polarized_xsections}).  The reason is that the cross
section for the charged pair production consists not only of
the ISR but also of the FSR parts which have different $x_\gamma$-dependent
radiator functions as well as slightly different $x_\gamma$-dependent polarization
factors. Note that the initial polarization factor is a function of $q^2$, i.e.
$x_\gamma$, while the final polarization factor is constant for a given
$\sqrt{s}$. Neglecting the slight variations due to the FSR contributions,
the LR ratio ${\cal R}_{LR}$ is given to a good approximation by
\begin{eqnarray}
        {\cal R}_{LR} [X]
\simeq  \left\{\begin{array}{cl}
         4 s^4_W \approx 0.199   &
         \ \ \mbox{for}\quad X = \chi^-_H, \tilde{\ell}^-_L \\[2mm]
         0         & \ \ \mbox{for}\quad X = \chi^-_W
         \end{array}\right.
\end{eqnarray}
The approximately zero ratio ${\cal R}_{LR}$ in the wino-type chargino case can be
traced to the perfect cancellation between the $\gamma$ and $Z$ exchange
diagrams for the right-handed electron beam polarization,
i.e. $1+c_R\, c^Z_{\chi^-_W} = 1-(s_W/c_W)
(c_W/s_W)=0$ in the asymptotic limit.

The right frame of Fig.$\,$\ref{fig:ratio_of_polarized_xsections} shows the
LR ratio of the inclusive sum of the charged and neutral pair production cross
sections in each scenario. Again this inclusive LR ratio remains almost constant
and enables us to distinguish the $W_{1/2}$ scenario from the $H_{1/2}$ and
$L_0$ scenarios.

\subsection{Alternative discrimination methods}
\label{subsec:alternative_methods}

While the Higgsino-type neutralino $\chi^0_H$ in the $H_{1/2}$
scenario is a Dirac fermion, the wino-type neutralino $\chi^0_W$ in the $W_{1/2}$
scenario is a Majorana fermion. Unlike the Dirac neutralino the Majorana
neutralino $\chi^0_W$ can mediate via a $t$-channel exchange a typical
fermion-number violating process such as the same-sign chargino-pair production
process, $W^-W^-\to\chi^-_W \chi^-_W$. The possible $e^-e^-$ collision mode of
the ILC experiments enables us to distinguish the $W_{1/2}$ scenario from
the $H_{1/2}$ scenario by searching for the same-sign $WW$ fusion process via
the process $e^-e^-\to \nu_e\nu_e W^- W^- \to \nu_e\nu_e \chi^-_W \chi^-_W$.

Although the neutral state $X^0$ of the (nearly) degenerate pair $[X^-,X^0]$ in each
scenario is stable, the charged state $X^-$ can decay to $X^0$ via charged current
interactions. For the typical loop-induced mass differences of a few hundred
MeV the most important decay modes are $X^-\to X^0\pi^-, X^0 e^- \bar{\nu}_e$ and
$X^0 \mu^-\bar{\nu}_\mu$. The decay products typically have low $p_T$, but as
demonstrated for the proposed International Large Detector (ILD) at the ILC tracking
efficiency of $60\%$ can be expected down to $p_T$ values of
200 MeV \cite{Behnke:2013lya}. On the other hand, the inner layer of the ILD
vertex detector would be extended down to the radius of 1.6 cm, therefore
offering good prospects of observing $X^-$ tracks, which in this case would have
a decay length of ${\cal O}(10\, {\rm cm})$ or less. The combination of different
detection methods based on the massive charged tracks, displaced vertices, and soft
decay products will enable us to cover all mass differences. Since relatively low data
volumes are expected, no hardware trigger would be needed allowing for search of
rare processes. Even in the case when the decays products can be observed, all
scenarios analyzed here would lead to the same final state. The angular photon
distributions would therefore offer a convenient discrimination method.
Finally, angular distributions of the decay products would provide additional
information on the spin, but such an analysis is beyond scope of the present study.

\section{Summary and conclusions}
\label{sec:summary_conclusions}

Given the current null results for SUSY searches at the LHC, we were strongly
motivated to consider the situation in which the only accessible SUSY states are
EW gauginos, Higgsinos or sleptons.
We explored three characteristic scenarios, each of which has a nearly
degenerate pair of a charged state and a neutral state with a small mass
difference. In the framework of MSSM  the three cases can be characterized as
(a) two spin-1/2 Higgsino SU(2)$_L$  doublets, (b) a spin-1/2 wino
SU(2)$_L$ triplet and (c) a spin-0 left-handed slepton SU(2)$_L$ doublet beyond the
SM particle spectrum. We presented the theoretical structures, their
interactions with the SM fields and their radiatively-induced mass splitting
in Sec.$\,$\ref{sec:scenarios}.

Due to near mass degeneracy, not only the neutral particle but also the charged
particle of each pair is not easily detectable in collider experiments.
We first presented the analytical expressions for the pair production of an
invisible neutral pair involving a hard photon emission, and discussed their
general features from the initial state radiation (ISR) and the final state
radiation (FSR) in Sec.$\,$\ref{sec:single-photon_processes}.
In our numerical studies, we illustrated our results with a 500 GeV ILC.
We provided a detailed and systematic analysis
with polarized electron and positron polarizations so as
to check the detectability of the charged particles as well as neutral particles
and how well their properties can be characterized.
As discussed in Sec.~\ref{subsec:isr_fsr_comparison}, the FSR effect in the
spin-1/2 charged pair production, compared to the ISR part, decreases monotonically
in size from about $40\ (10)\,\%$ for $x_\gamma=0.04$ and becomes
negligible close to the threshold with $x_\gamma=0.84\ (0.36)$ for
$m_X=100\ (200)$ GeV and $\sqrt{s}=500$ GeV. Therefore, the previous analyses
in the literature based on the ISR approximation are rather reliable,
especially when the mass $m_X$ is not
far from the half of the $e^+e^-$ CM energy $\sqrt{s}$. On the contrary,
in the spin-0 charged pair production, the FSR effect becomes larger than the ISR
part near the threshold as shown in Fig.$\,$\ref{fig:ratio_fsr_to_isr}, which
might endanger any consequences based simply on the ISR approximation.
Nevertheless, we found that, in spite of the FSR contamination,
the results based on the ISR approximation are quantitatively very similar to those
with both the ISR and FSR parts.

In Sec.$\,$\ref{subsec:significance}, we studied the signal observation with
respect to the SM backgrounds. We also demonstrated in
Sec.$\,$\ref{subsec:spin_determination} that
the excitation pattern near the threshold can be exploited through the photon
energy distribution to determine the spin of
the SUSY EW particles unambiguously. The (normalized) photon energy distribution
near threshold shows a steeply-rising $S$-wave excitation for a spin-1/2 pair
while a slowly-rising $P$-wave excitation for a spin-0 pair, even after
the contamination from the FSR part is included
(see Fig.$\,$\ref{fig:threshold_behavior}).

Furthermore, the LR ratio of right-handed and left-handed cross sections introduced
in Sec.$\,$\ref{subsec:ratios} takes very different values according to the production
modes; $\sim s^4_W/(1/2-s^2_W)^2\simeq 0.65$ for $X=\chi^0_D$ and
$\tilde{\nu}_\ell$; $\sim 4 s^4_W\simeq 0.20$ for $X=\chi^-_H$ and
$\tilde{\ell}^-_L$; and $\sim 0$ for $X=\chi^-_W$. Even after taking
the inclusive sum of the charged and neutral modes in each scenario, the LR ratio
has a nearly constant value unique to the scenario of the lower-lying
EW particles as shown in Fig.$\,$\ref{fig:ratio_of_polarized_xsections}.
Therefore, in addition to enhancing the statistical significance sizably,
the electron and positron beam
polarizations are very powerful in characterizing the production modes.
Combining the LR ratio and the threshold excitation pattern, we can identify
unambiguously which scenario among the three scenarios is realized.
Our analyses are easily generalizable to other collider energies as long as
the pair production is kinematically accessible.

Our analytic and numerical results demonstrate clearly
the strong physics potential of the ILC in detecting and characterizing the
invisible particles, complementary to the very difficult searching environment
at the LHC. Further detailed analyses and detector simulations may be needed to
reach fully realistic conclusions at the ILC.

\vskip 0.5cm

\noindent
{\bf Acknowledgements.} The work of SYC was supported in part by Basic
Science Research Program through the National Research Foundation (NRF) funded
by the Ministry of Education, Science and Technology (NRF-2011-0010835) and
in part by research funds of Chonbuk National University in 2013.
TH and XW were supported in part by the U.S.~Department of Energy under grant
No.~DE-FG02-95ER40896, in part by the PITT PACC.
Work was partially supported by the Polish National Science Centre under
research grants  DEC-2012/05/B/ST2/02597 and
OPUS-2012/05/B/ST2/03306. KR has been partially supported by the MINECO, Spain, under contract FPA2013-44773-P;
Consolider-Ingenio CPAN CSD2007-00042; Spanish MINECO Centro de excelencia Severo Ochoa Program under grant SEV-2012-0249 and by JAE-Doc programme.


\vskip 1.5cm

\begin{thebibliography}{99}

\bibitem{Aad:2012tfa}
  G.~Aad {\it et al.}  [ATLAS Collaboration],
  Phys.\ Lett.\ B {\bf 716}, 1 (2012)  [arXiv:1207.7214 [hep-ex]].

\bibitem{Chatrchyan:2012ufa}
  S.~Chatrchyan {\it et al.}  [CMS Collaboration],
  Phys.\ Lett.\ B {\bf 716}, 30 (2012)  [arXiv:1207.7235 [hep-ex]].

\bibitem{mass125}
  M.~Duehrssen, talk and Moriond 2015,
  https://indico.in2p3.fr/event/10819/session/3/contribution/102/material/slides/1.pdf.

\bibitem{Caola:2013yja}
  F.~Caola and K.~Melnikov,
  Phys.\ Rev.\ D {\bf 88}, 054024 (2013)
  [arXiv:1307.4935 [hep-ph]].

\bibitem{Khachatryan:2014iha}
  V.~Khachatryan {\it et al.}  [CMS Collaboration],
  Phys.\ Lett.\ B {\bf 736}, 64 (2014)
  [arXiv:1405.3455 [hep-ex]].

\bibitem{Wilson:1970ag}
  K.~G.~Wilson,
  Phys.\ Rev.\ D {\bf 3}, 1818 (1971).

\bibitem{Gildener:1976ai}
  E.~Gildener,
  Phys.\ Rev.\ D {\bf 14}, 1667 (1976).

\bibitem{Weinberg:1978ym}
  S.~Weinberg,
  Phys.\ Lett.\ B {\bf 82}, 387 (1979).

\bibitem{'tHooft:1979bh}
  G.~'t Hooft,
  NATO Sci.\ Ser.\ B {\bf 59}, 135 (1980).

\bibitem{Nilles:1983ge}
  H.~P.~Nilles,
  Phys.\ Rept.\  {\bf 110}, 1 (1984).

\bibitem{Haber:1984rc}
  H.~E.~Haber and G.~L.~Kane,
  Phys.\ Rept.\  {\bf 117}, 75 (1985).

\bibitem{Griest:2000kj}
  K.~Griest and M.~Kamionkowski,
  Phys.\ Rept.\  {\bf 333}, 167 (2000).

\bibitem{Bertone:2004pz}
  For a recent review on particle dark matter, see {\it e.g.~}
  G.~Bertone, D.~Hooper and J.~Silk,
  Phys.\ Rept.\  {\bf 405}, 279 (2005)  [hep-ph/0404175].

\bibitem{Giudice:2004tc}
  G.~F.~Giudice and A.~Romanino,
  Nucl.\ Phys.\ B {\bf 699}, 65 (2004)
  [Erratum-ibid.\ B {\bf 706}, 65 (2005)]
  [hep-ph/0406088].

\bibitem{ArkaniHamed:2004yi}
  N.~Arkani-Hamed, S.~Dimopoulos, G.~F.~Giudice and A.~Romanino,
  Nucl.\ Phys.\ B {\bf 709}, 3 (2005)
  [hep-ph/0409232].

\bibitem{Wells:2004di}
  J.~D.~Wells,
  Phys.\ Rev.\ D {\bf 71}, 015013 (2005)
  [hep-ph/0411041].

\bibitem{Baer:2012uy}
  H.~Baer, V.~Barger, P.~Huang and X.~Tata,
  JHEP {\bf 1205}, 109 (2012)
  [arXiv:1203.5539 [hep-ph]].

\bibitem{Papucci:2011wy}
  M.~Papucci, J.~T.~Ruderman and A.~Weiler,
  JHEP {\bf 1209}, 035 (2012)
  [arXiv:1110.6926 [hep-ph]].

\bibitem{Jung:2013zya}
  S.~Jung and J.~D.~Wells,
  Phys.\ Rev.\ D {\bf 89}, no. 7, 075004 (2014)
  [arXiv:1312.1802 [hep-ph]].

\bibitem{Chen:1995yu}
  C.~H.~Chen, M.~Drees and J.~F.~Gunion,
  Phys.\ Rev.\ Lett.\  {\bf 76}, 2002 (1996)  [hep-ph/9512230].

\bibitem{Chen:1996ap}
  C.~H.~Chen, M.~Drees and J.~F.~Gunion,
  Phys.\ Rev.\ D {\bf 55}, 330 (1997)  [Erratum-ibid.\ D {\bf 60}, 039901 (1999)]
  [hep-ph/9607421].

\bibitem{Giudice:2010wb}
  G.~F.~Giudice, T.~Han, K.~Wang and L.~-T.~Wang,
  Phys.\ Rev.\ D {\bf 81}, 115011 (2010)  [arXiv:1004.4902 [hep-ph]].

\bibitem{Han:2013kza}
  T.~Han, S.~Padhi and S.~Su,
  Phys.\ Rev.\ D {\bf 88}, 115010 (2013)  [arXiv:1309.5966 [hep-ph]].

\bibitem{Cheung:2005pv}
  K.~Cheung, C.~-W.~Chiang and J.~Song,
  JHEP {\bf 0604}, 047 (2006)  [hep-ph/0512192].

\bibitem{Baer:2011ec}
  H.~Baer, V.~Barger and P.~Huang,
  JHEP {\bf 1111}, 031 (2011)  [arXiv:1107.5581 [hep-ph]].

\bibitem{Hensel:2002bu}
  C.~Hensel,
  DESY-THESIS-2002-047.

\bibitem{Berggren:2013vfa}
  M.~Berggren, F.~Br\"ummer, J.~List, G.~Moortgat-Pick, T.~Robens, K.~Rolbiecki and H.~Sert,
  Eur.\ Phys.\ J.\ C {\bf 73}, 2660 (2013)  [arXiv:1307.3566 [hep-ph]].

\bibitem{Cheng:1998hc}
  H.~-C.~Cheng, B.~A.~Dobrescu and K.~T.~Matchev,
  Nucl.\ Phys.\ B {\bf 543}, 47 (1999)  [hep-ph/9811316].

\bibitem{Gherghetta:1999sw}
  T.~Gherghetta, G.~F.~Giudice and J.~D.~Wells,
  Nucl.\ Phys.\ B {\bf 559}, 27 (1999)  [hep-ph/9904378].

\bibitem{Cheung:2005ba}
  K.~Cheung and C.~-W.~Chiang,
  Phys.\ Rev.\ D {\bf 71}, 095003 (2005)  [hep-ph/0501265].

\bibitem{Feng:1999fu}
  J.~L.~Feng, T.~Moroi, L.~Randall, M.~Strassler and S.~-f.~Su,
  Phys.\ Rev.\ Lett.\  {\bf 83}, 1731 (1999)  [hep-ph/9904250].

\bibitem{Ibe:2006de}
  M.~Ibe, T.~Moroi and T.~T.~Yanagida,
  Phys.\ Lett.\ B {\bf 644}, 355 (2007)  [hep-ph/0610277].

\bibitem{Dimopoulos:1990kc}
  S.~Dimopoulos, N.~Tetradis, R.~Esmailzadeh and L.~J.~Hall,
  Nucl.\ Phys.\ B {\bf 349}, 714 (1991)  [Erratum-ibid.\ B {\bf 357}, 308 (1991)].

\bibitem{Thomas:1998wy}
  S.~D.~Thomas and J.~D.~Wells,
  Phys.\ Rev.\ Lett.\  {\bf 81}, 34 (1998)  [hep-ph/9804359].

\bibitem{Sher:1995tc}
  M.~Sher,
  Phys.\ Rev.\ D {\bf 52}, 3136 (1995)  [hep-ph/9504257].

\bibitem{Behnke:2013xla}
  T.~Behnke, J.~E.~Brau, B.~Foster, J.~Fuster, M.~Harrison, J.~M.~Paterson, M.~Peskin
  and M.~Stanitzki {\it et al.},
  arXiv:1306.6327 [physics.acc-ph].

\bibitem{Baer:2013cma}
  H.~Baer, T.~Barklow, K.~Fujii, Y.~Gao, A.~Hoang, S.~Kanemura, J.~List and H.~E.~Logan
  {\it et al.},
  arXiv:1306.6352 [hep-ph].

\bibitem{Behnke:2013lya}
  T.~Behnke, J.~E.~Brau, P.~N.~Burrows, J.~Fuster, M.~Peskin, M.~Stanitzki, Y.~Sugimoto and
  S.~Yamada {\it et al.},
  arXiv:1306.6329 [physics.ins-det].

\bibitem{Chatrchyan:2012me}
  S.~Chatrchyan {\it et al.}  [CMS Collaboration],
  JHEP {\bf 1209}, 094 (2012)
  [arXiv:1206.5663 [hep-ex]].

\bibitem{ATLAS:2012ky}
  G.~Aad {\it et al.}  [ATLAS Collaboration],
  JHEP {\bf 1304}, 075 (2013)
  [arXiv:1210.4491 [hep-ex]].

\bibitem{Birkedal:2004xn}
  A.~Birkedal, K.~Matchev and M.~Perelstein,
  Phys.\ Rev.\ D {\bf 70}, 077701 (2004)
  [hep-ph/0403004].

\bibitem{Ma:1978zm}
  E.~Ma and J.~Okada,
  Phys.\ Rev.\ Lett.\  {\bf 41}, 287 (1978)  [Erratum-ibid.\  {\bf 41}, 1759 (1978)].

\bibitem{Gaemers:1978fe}
  K.~J.~F.~Gaemers, R.~Gastmans and F.~M.~Renard,
  Phys.\ Rev.\ D {\bf 19}, 1605 (1979).

\bibitem{Barbiellini:1981zm}
  G.~Barbiellini, B.~Richter and J.~Siegrist,
  Phys.\ Lett.\ B {\bf 106}, 414 (1981).

\bibitem{Grassie:1983kq}
  K.~Grassie and P.~N.~Pandita,
  Phys.\ Rev.\ D {\bf 30}, 22 (1984).

\bibitem{Fayet:1986zc}
  P.~Fayet,
  Phys.\ Lett.\ B {\bf 175}, 471 (1986).

\bibitem{Dicus:1990vm}
  D.~A.~Dicus, S.~Nandi and J.~Woodside,
  Phys.\ Lett.\ B {\bf 258}, 231 (1991).

\bibitem{Lopez:1996gd}
  J.~L.~Lopez, D.~V.~Nanopoulos and A.~Zichichi,
  Phys.\ Rev.\ Lett.\  {\bf 77}, 5168 (1996)  [hep-ph/9609524].

\bibitem{Dreiner:2006sb}
  H.~K.~Dreiner, O.~Kittel and U.~Langenfeld,
  Phys.\ Rev.\ D {\bf 74}, 115010 (2006)  [hep-ph/0610020].

\bibitem{Choi:1999bs}
  S.~Y.~Choi, J.~S.~Shim, H.~S.~Song, J.~Song and C.~Yu,
  Phys.\ Rev.\ D {\bf 60}, 013007 (1999)  [hep-ph/9901368].


\bibitem{Bartels:2012ex}
  C.~Bartels, M.~Berggren and J.~List,
  Eur.\ Phys.\ J.\ C {\bf 72}, 2213 (2012)
  [arXiv:1206.6639 [hep-ex]].

\bibitem{Pierce:1996zz}
  D.~M.~Pierce, J.~A.~Bagger, K.~T.~Matchev and R.~-j.~Zhang,
  Nucl.\ Phys.\ B {\bf 491}, 3 (1997)  [hep-ph/9606211].

\bibitem{Nicrosini:1989pn}
  O.~Nicrosini and L.~Trentadue,
  Phys.\ Lett.\ B {\bf 231}, 487 (1989).

\bibitem{Montagna:1995wp}
  G.~Montagna, O.~Nicrosini, F.~Piccinini and L.~Trentadue,
  Nucl.\ Phys.\ B {\bf 452}, 161 (1995)  [hep-ph/9506258].

\bibitem{Berends:1984dw}
  F.~A.~Berends and R.~Kleiss,
  Nucl.\ Phys.\ B {\bf 260}, 32 (1985).

\bibitem{Dokshitzer:1994jt}
  Y.~L.~Dokshitzer, V.~A.~Khoze and W.~J.~Stirling,
  Nucl.\ Phys.\ B {\bf 428}, 3 (1994)  [hep-ph/9405243].

\bibitem{Low:1958sn}
  F.~E.~Low,
  Phys.\ Rev.\  {\bf 110}, 974 (1958).

\bibitem{Burnett:1967km}
  T.~H.~Burnett and N.~M.~Kroll,
  Phys.\ Rev.\ Lett.\  {\bf 20}, 86 (1968).

\bibitem{Agashe:2014kda}
  K.~A.~Olive {\it et al.}  [Particle Data Group Collaboration],
  Chin.\ Phys.\ C {\bf 38}, 090001 (2014).


\end{thebibliography}
\end{document}